\documentclass[RNAAS]{aastex62}
\usepackage[utf8]{inputenc}
\usepackage[english]{babel}
\usepackage{amsmath}
\usepackage{amsfonts}
\usepackage{amssymb}
\usepackage{graphicx}

\begin{document}

\title{KiDS0239-3211: A new gravitational quadruple lens candidate}
\correspondingauthor{Alexey Sergeyev}
\email{alexey.v.sergeyev@gmail.com}

\author{A. Sergeyev}
\affiliation{V.N. Karazin Kharkiv National University, Ukraine}
\affiliation{Institute of Radio Astronomy of the National Academy of Sciences of Ukraine}
\author{C. Spiniello}
\affiliation{INAF - Osservatorio Astronomico di Capodimonte, Naples, Italy} 
\affiliation{European Southern Observatory, Garching, Germany} 
\author{V. Khramtsov}
\affiliation{V.N. Karazin Kharkiv National University, Ukraine}\affiliation{Department of Astronomy and Space Informatics V.N. Karazin Kharkiv National University Kharkiv, Ukraine}
\author{N. R. Napolitano}
\affiliation{INAF - Osservatorio Astronomico di Capodimonte, Naples, Italy} 
\author{E. Bannikova}
\affiliation{V.N. Karazin Kharkiv National University, Ukraine}
\author{C. Tortora}
\affiliation{INAF-Osservatorio Astrofisico di Arcetri, Firenze, Italy}
\author{F. I. Getman}
\affiliation{INAF - Osservatorio Astronomico di Capodimonte, Naples, Italy} 
\author{A. Agnello}
\affiliation{European Southern Observatory, Garching, Germany} 






\section{Introduction} 
Strongly lensed quasars (SLQSOs), particularly quadruply lensed systems, are very rare (\citealt{Oguri10}), but very valuable probes of cosmology and extragalactic astrophysics. 

With the KiDS Strongly lensed QUAsar Detection project (KiDS-SQuaD, \citealt{Spiniello18}, hereafter S18) we have started a systematic census of lensed quasars in the Kilo Degree Survey (KiDS, \citealt{deJong15}), taking advantage from the
high spatial resolution of VST (0.2$\arcsec$/pixel) and its stringent seeing constraints ($<0.8\arcsec$ in r-band). 

In S18 we selected qso-like objects  based on infrared and optical colors and then used different methods to identify multiple systems, that we then visually inspected. 
However, with these methods, the number of candidates highly depends on the (somehow arbitrary and often calibrated on previous finding) selection criteria.  Generally this number is of the order of thousands every 100 deg$^2$, making the visual inspection long and tedious. 
Thus, to make our research more effective and suitable to deal with larger amount of data coming with future data releases and new wide-sky surveys (e.g. Euclid or LSST), we developed a new method based on machine learning. 
These techniques have the great advantage to explore, with little computing time, large amount of candidates with less stringent pre-selections and with reasonably high precision (recovery rate) and little contamination (spurious detection). 
These have been applied already for the search of strong gravitational arcs in KiDS (\citealt{Petrillo18}) and SLQSOs in other wide-sky Surveys (\citealt{Agnello15}). 

\section{Mining Strong lensed quasars with machine learning}
Although the combination of near-infrared and optical colours is 
the most effective way to separate quasars from stars within photometric surveys (\citealt{Akhmetov17}),
infrared information is not always available and often not deep enough. 
For this reason, we developed a machine learning based method to separate QSOs from stars using only 4 photometric optical bands (u,g,r,i).  We used Random Forests classifier (self-written python code) with spectroscopically confirmed stars ($622,052$) and QSOs ($484,372$) from SDSS DR14 (\citealt{Abolfathi18}). 
We worked in the 6-dimensional colour space, made of all possible differences of magnitudes in the different bands. 
The classification was performed with a 5-fold cross-validation, obtaining 99\% of purity and 65\% of completeness of the quasar validating sample. 
We thus apply this procedure on the KiDS-DR3 catalog and then select QSO-like multiple sources (with a separation $\le 10\arcsec$). We ended up with 3187 candidates, for which we inspected the combined ugr KiDS cutouts. A detailed presentation of the method will be presented in a future publication. 

\pagebreak
\section{KiDS 0239-3211}
KiDS 0239-3211 (Ra: $02$:$39$:$29.69$, Dec: $-32$:$11$:$29.66$) is without doubt the most promising candidate we found in the public KiDS-DR3. 
\begin{figure}[h!]
\begin{center}
\includegraphics[scale=0.53,angle=0]{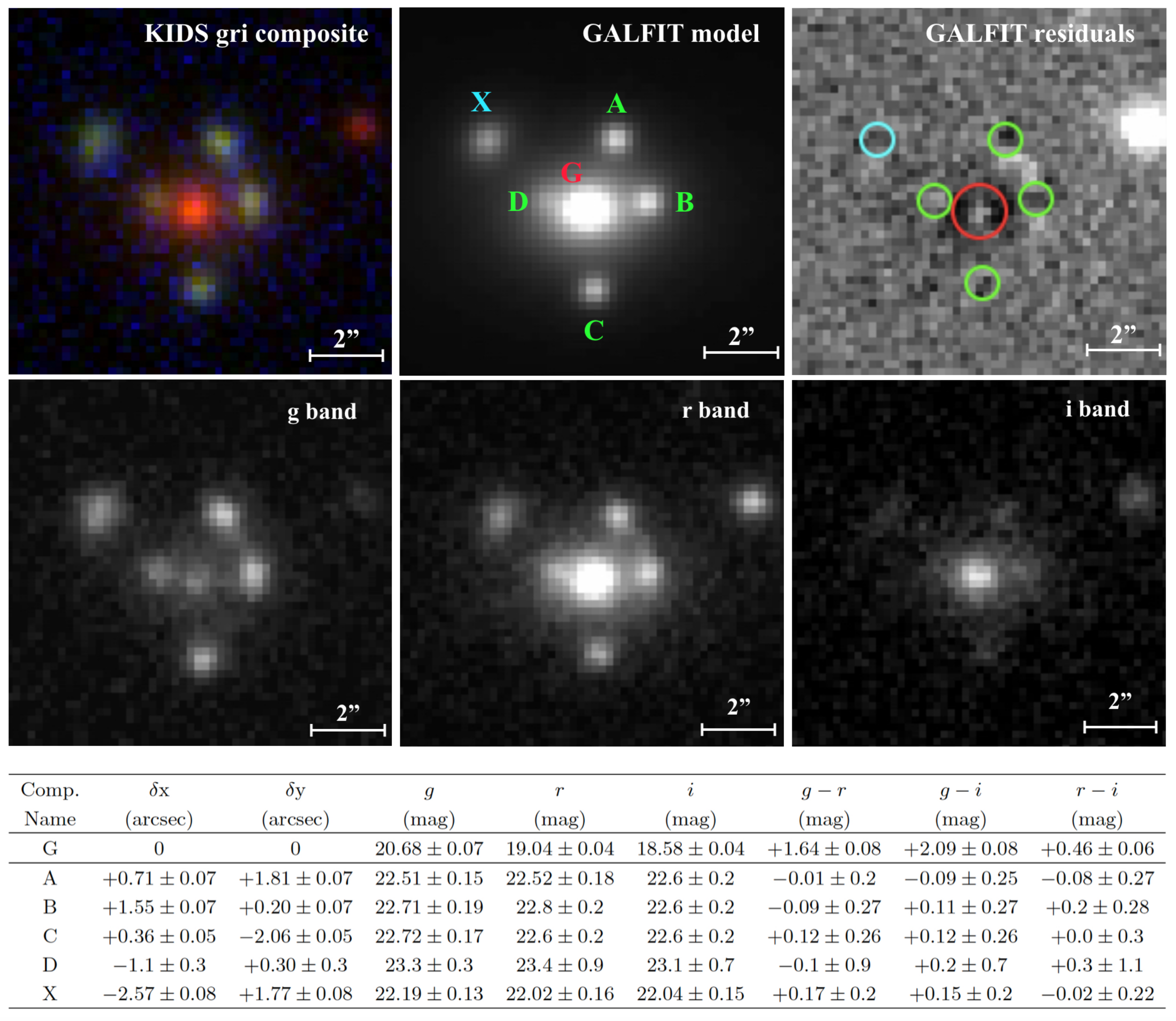}
\end{center}
\end{figure}

In the figure we show the combined $gri$ KiDS cutout of the quad, as well as single band images and GALFIT model and residuals. The system is composed by a red galaxy in the middle (G) surrounded by five blue blobs within few arcsec distance (listed in the Table)
We obtained differential photometry directly from GALFIT, calibrating the zero-point with a reference star from the KiDS catalog. We note that all blobs have consistent colors, within the errors, which support the lensing hypothesis.

Given the geometry, colors and shape of the blobs, we believe that four of them are multiple images of a point-like source (A, B, C, D in green) whereas the fifth, more distant one is a contaminant (X in cyan). This hypothesis is further confirmed by the machine learning based estimate of the photometric redshifts ($z\sim0.5$ for G and X and $z\sim1.6-2.0$ for A and C, from the catalog in \citealt{deJong17} and derived using the Multi Layer Perceptron with Quasi Newton Algorithm  technique presented in \citealt{Cavuoti15}).
\pagebreak

\section{Discussion and conclusions} 
The discovery of KiDS 0239-3211, missed by other methods, demonstrates that our machine learning set-up, although still preliminary, represents a considerable step further to effectively find new SLQSOs in present and future public wide-sky photometric surveys. 

Nevertheless, to unambiguously confirm the lensing nature of the system and to translate the `geometric' lens model results (e.g. Einstein radius) into physical measurements of luminous and dark masses, the redshifts of the source and of the deflector are needed. 
We are therefore setting up a systematic, multi-site, multi-facility campaign for spectroscopic observation of the best candidates in KiDS, including KiDS 0239-3211, but the publication of the object coordinates and fluxes is meant to encourage the community to help out with this task.

\section{Acknowledgments} 
We thank the organizators and participants of the Italy-Ukraine Collaboration in Astronomy Meeting, during which this discovery was made. We acknowledge the KiDS collaboration for the work done to make the data publicly available. 
CS and NRN have received funding from the European Union's Horizon 2020 research and innovation programme. under the Marie Sk\l odowska-Curie actions grant agreements No 664931 and No 721463 to the SUNDIAL ITN network.


\begin{thebibliography}{20}
\bibitem[Abolfathi et al.(2018)]{Abolfathi18} Abolfathi, B., Aguado, D.~S., Aguilar, G., et al.\ 2018, \apjs, 235, 42 
\bibitem[Akhmetov et al.(2017)]{Akhmetov17} Akhmetov V.~S., Fedorov P.~N., Velichko A.~B., Shulga V.~M., 2017, \mnras, 469, 763 
\bibitem[Agnello et al.(2015)]{Agnello15} Agnello, A., Kelly, B.~C., Treu, T., \& Marshall, P.~J.\ 2015, \mnras, 448, 1446 
\bibitem[Cavuoti et al.(2015)]{Cavuoti15} Cavuoti et al. 2015, \mnras, 452, 3100
\bibitem[de Jong et al.(2015)]{deJong15} de Jong J. T. A., Verdoes Kleijn, G.~A., Boxhoorn, D.~R., et al., 2015, A\&A, 582, A62
\bibitem[de Jong et al.(2017)]{deJong17} de Jong, J.~T.~A., Verdois Kleijn, G.~A., Erben, T., et al.\ 2017, \aap, 604, A134 
\bibitem[Oguri \& Marshall(2010)]{Oguri10} Oguri, M., \& Marshall, P.~J.\ 2010, \mnras, 405, 2579 
\bibitem[Petrillo et al.(2018)]{Petrillo18} Petrillo, C.~E., Tortora, C., Chatterjee, S., et al.\ 2018, arXiv:1807.04764 
\bibitem[Spiniello et al.(2018)]{Spiniello18} Spiniello, C., Agnello, A., Napolitano, N.~R., et al.\ 2018, \mnras, 480, 1163 
\end{thebibliography}
\end{document}